\documentclass[journal,12pt,onecolumn,draftclsnofoot]{IEEEtran}

\usepackage{color}
\usepackage{multirow}
\usepackage{graphicx}
\usepackage{epstopdf}
\usepackage[cmex10]{amsmath}
\usepackage{amssymb}

\usepackage{multirow}
\usepackage{amsthm}
\usepackage{cite}
\usepackage{acronym} 
\usepackage{algorithm, algcompatible}

\algnewcommand\INPUT{\item[\textbf{Input:}]}
\algnewcommand\OUTPUT{\item[\textbf{Output:}]}

\allowdisplaybreaks

\begin{document}

\title{Non-Linear Age of Information: An Energy Efficient Receiver-Centric Approach} 
\author{Nikolaos~I.~Miridakis, Theodoros~A.~Tsiftsis and Guanghua~Yang
\thanks{\textit{Corresponding Author: T. A. Tsiftsis.}}
\thanks{The authors are with the Institute of Physical Internet and School of Intelligent Systems Science \& Engineering, Jinan University, Zhuhai Campus, Zhuhai 519070, China. N. I. Miridakis is also with the Dept. of Informatics and Computer Engineering, University of West Attica, Aegaleo 12243, Greece (e-mails: nikozm@uniwa.gr, theo\_tsiftsis@jnu.edu.cn, ghyang@jnu.edu.cn).}
}


\maketitle

\begin{abstract}
The age of information (AoI) performance metric for point-to-point wireless communication systems is analytically studied under Rician-faded channels and when the receiver is equipped with multiple antennas. The general scenario of a non-linear AoI function is considered, which includes the conventional linear AoI as a special case. The stop-and-wait transmission policy is adopted, where the source node samples and then transmits new data only upon the successful reception of previous data. This approach can serve as a performance benchmark for any queuing system used in practice. New analytical and closed-form expressions are derived with respect to the average AoI and average peak AoI for the considered system configuration. We particularly focus on the energy efficiency of the said mode of operation, whereas some useful engineering insights are provided.
\end{abstract}


\IEEEpeerreviewmaketitle

\section{Introduction}
\IEEEPARstart{A}{ge} of information (AoI) denotes a rather new performance metric for wireless communication systems, which indicates the freshness of the latest received data. Doing so, it has numerous applications in a set of various recent and upcoming networking paradigms, e.g., Internet-of-things (IoT), vehicular and machine-type communications as well as low-latency and delay-critical services \cite{j:Yates2020}. On another front, source nodes in these applications are usually simple sensor devices which are battery-powered and thus become energy-constrained. Hence, keeping the status updates fresh (i.e., minimizing AoI) and providing energy efficiency at the same time represents quite a challenging and non-trivial task.

Typically, the penalty function used to describe AoI is a linear function of time. Nonetheless, the versatility of the underlying data exchanging process allows for a more general non-linear penalty function; the latter can suitably describe the \emph{value} of new information updates, which in turn may result to a non-linear AoI behavior. In fact, the proportion of non-linearity of the AoI penalty function indicates the importance of keeping updates as fresh as possible and to what extent affects the performance of a given application. In \cite{j:KostaPappas20}, the performance of a non-linear average AoI and its peak-average  (PAoI) counterpart were analytically studied under an M/M/1 queuing model. Also, a similar system model was analytically studied in \cite{j:ZhengZhouJiang19} when energy harvesting-enabled source sensor nodes are employed. However, these works assumed single-antenna transceivers, they did not consider possible user mobility/portability as well as their derivations were conditioned on a fixed packet error rate and/or service rate. Further, the energy/AoI tradeoff has been studied in \cite{c:GongChenMa18} and \cite{c:XieGongMa2020} for systems using retransmission policy and coded short-packet communication, respectively. Yet, their results were conditioned on a fixed/given packet error rate without modeling the nature of the wireless channel, the range of antenna array at the transceiver and the potential of user mobility. 

Capitalizing on the aforementioned observations, current work focuses on the non-linear AoI and PAoI of point-to-point wireless communication systems, where the source node may be either a static or mobile simple sensor monitoring device with an arbitrary moving speed profile. A single-antenna source node and a multiple-antenna destination are assumed. In addition, spatially independent Rician faded channels are considered; whereby non-line of sight (LoS), near-LoS and/or LoS signal propagation conditions can be effectively modeled. Since we focus on simple transmit devices with limited energy constraints, the stop-and-wait transmission policy is adopted herein, where the source does not sample and transmit new data until the previous message is successfully received. Also, the proposed approach focuses on the energy efficiency and analyzes the AoI/energy tradeoff in practical wireless communication setups. For the first time, new exact (unconditional in terms of the packet error rate) closed-form expressions are presented with respect to the introduced non-linear average AoI, PAoI as well as the energy efficiency under the considered wireless system configuration. 

{\it Notation:} $\mathbb{E}[\cdot]$ is the expectation operator; $\mathbb{E}_{x}[f(x)]$ denotes expectation of function $f(x)$ with respect to $x$; ${\rm Pr[\cdot]}$ stands for the probability operator; $y|z$ denotes that $y$ is conditioned on $z$ event; $I_{v}(\cdot)$ denotes the $v^{\rm th}$ order modified Bessel function of the first kind \cite[Eq. (8.445)]{tables}; $Q_{\nu}(\cdot,\cdot)$ is the generalized $\nu$th order Marcum-$Q$ function.

\section{System Model}
\begin{figure}[!t]
\centering
\includegraphics[trim=1.5cm 0.0cm 0.5cm .1cm, clip=true,totalheight=0.25\textheight]{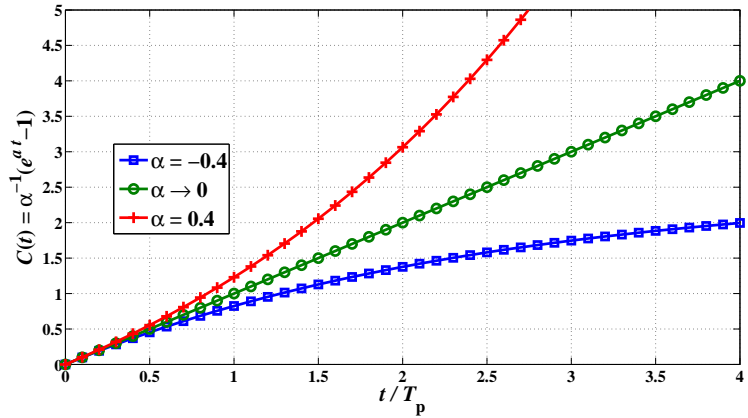}
\caption{Cost function with different $a$ values vs. normalized time.}
\label{fig1}
\end{figure}

Consider a single-antenna source node that generates status updates in the form of packets with a fixed length duration $T_{\rm p}$ at random intervals and sends them at a given destination equipped with $N$ antennas. {\color{black}Perfect channel state information is assumed at the receiver side; hence, maximum ratio combining (MRC) is performed so as to enhance the received channel gains.} To assess the freshness of the randomly generated updates, we expand the notion of the typical linear AoI to a generalized non-linear form via the following non-negative and monotonically increasing cost function:
\begin{align}
C(t)\triangleq a^{-1}\left(e^{a t}-1\right),\quad a\neq 0,
\label{costfunc}
\end{align} 
where $a\in \mathbb{R}$ is an application-dependent tuning parameter. It is noted that $C(t)\rightarrow t$ as $a\rightarrow 0$ (i.e., the conventional linear AoI); it behaves as an exponential-like function for $a>0$; while it resembles a logarithmic-like function for $a<0$. In Fig.~\ref{fig1}, the considered cost function is plotted for different $a$ parameter values. Consequently, the non-linear AoI is defined as $C(t-t_{\rm c})$, where $t-t_{\rm c}$ indicates the time elapsed from the last correctly received message (occurred at time $t_{\rm c}$). Typically, the mutual information of the data generation random process can be used to quantify the $a$ parameter; namely $\mathcal{I}(s_{t};s_{t-\tau})$ where $s_{t}$ denotes the source signal at time $t$ and $\tau$ is a certain time lag. Illustrative examples of $\mathcal{I}(\cdot;\cdot)$ for popular distribution models can be found in \cite{c:SunCyr18}. Therefore, as the mutual information decreases, the source process is less-correlated and an exponential-like cost function $C(t)$ seems more appropriate to characterize the penalty on AoI; in this case $a>0$. Quite the opposite holds for high-correlated source functions, which can be modeled by a log-like $C(t)$ by setting $a<0$. For moderate mutual information values, $a\rightarrow 0$ defines the conventional linear AoI.\footnote{Overall, we stress that the suitable tuning of $a$ parameter is in principle application-dependent. For instance, if the underlying process is the estimation of instantaneous wireless channel status, keeping the updates as fresh as possible is the prime goal regardless of the amount of time correlation between consecutive samples.} This is a reasonable assumption and also in line with other studies, e.g., \cite{j:KostaPappas20}. Alternatively, the autocorrelation operator with the said time lag $\tau$ can be applied to specify $a$ (e.g., if the distribution of data process is unknown or does not exist). Proper values of $\tau$ are further discussed in the next section.    

\begin{figure}[!t]
\centering
\includegraphics[trim=.9cm 0.5cm 1.0cm 1.0cm, clip=true,totalheight=0.25\textheight]{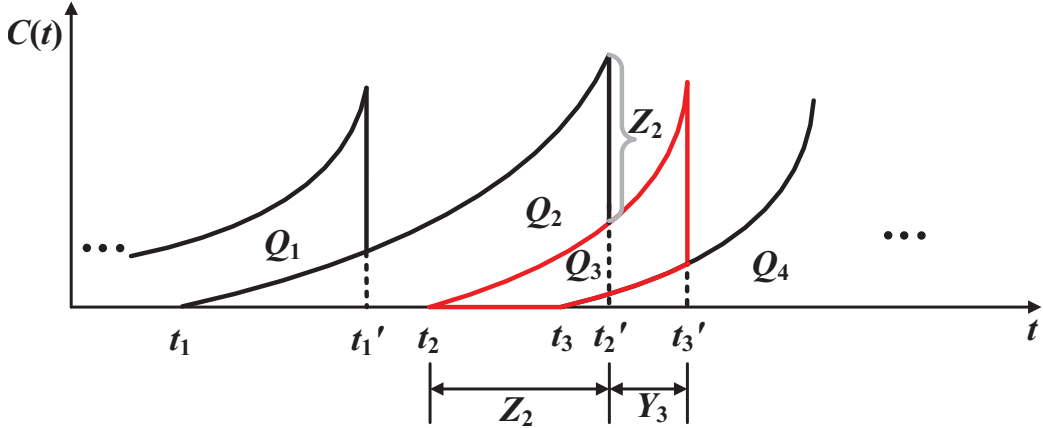}
\caption{An example of the considered non-linear AoI model where $a>0$.}
\label{fig2}
\end{figure}

Let $t_{i}$ and $t'_{i}$ be the time instance of the $i^{\rm th}$ packet generation and error-free reception, respectively, with $t_{i}<t'_{i}$. Also, let $Y_{i}\triangleq t'_{i}-t'_{i-1}$ and $Z_{i}\triangleq t'_{i}-t_{i}$ be the interarrival time between two consecutive (independent) successful packet receptions and transmission time duration of the $i^{\rm th}$ successfully received packet, respectively. The considered non-linear AoI model is illustrated in Fig.~\ref{fig2}. In the case when a transmission failure of a certain packet occurs, the source retransmits the same packet until a successful reception or when a maximum number of retransmissions, denoted as $M$, is reached. If the current transmission fails when $M$ retransmissions are reached, a new packet is generated and transmitted while AoI still increases until a successful reception occurs. As status sensing/sampling process consumes energy, a retransmission policy reflects on energy-efficient communications (since only the amount of energy used for packet transmission is required for each retransmission). On the other hand, a higher $M$ means a potentially longer AoI, introducing an AoI-energy efficiency tradeoff.  

\section{Performance Metrics}
For a \emph{stationary} and \emph{ergodic} information updating process, the time average AoI tends to its ensemble average counterpart given an asymptotically high number of consecutive samples. To this end, the average AoI reads as
\begin{align}
\overline{C}\triangleq \lim_{k\rightarrow+\infty}\frac{\sum^{k}_{i=1}Q_{i}}{\sum^{k}_{i=1}Y_{i}}=\frac{\mathbb{E}[Q]}{\mathbb{E}[Y]},
\label{avC}
\end{align}
where $\mathbb{E}[Q]\triangleq(1/k)\sum^{k}_{i=1}Q_{i}$ and $\mathbb{E}[Y]\triangleq(1/k)\sum^{k}_{i=1}Y_{i}$ (with $k\rightarrow +\infty$). Also, $Q_{i}$ denotes the disjoint area in Fig.~\ref{fig2} (e.g., $Q_{3}$ is the area within the annulus sector highlighted in red color), which is computed as
\begin{align}
\nonumber
Q_{i}&=a^{-1}\left[\int^{t'_{i}}_{t_{i-1}}\left(e^{a (t-t_{i-1})}-1\right)dt-\int^{t'_{i}}_{t_{i}}\left(e^{a (t-t_{i})}-1\right)dt\right]\\
&=\frac{1}{a^{2}}\left[e^{a\left(Y_{i}+Z_{i-1}\right)}-e^{a Z_{i}}\right]-\frac{1}{a}\left[Z_{i-1}+Y_{i}-Z_{i}\right].
\label{Qi}
\end{align}
In order to further proceed, the auxiliary integer-valued RVs $\tilde{Y}_{i}$ and $\tilde{Z}_{i}$ are introduced, such that $Y_{i}\triangleq T_{\rm p}\tilde{Y}_{i}$ and $Z_{i}\triangleq T_{\rm p}\tilde{Z}_{i}$ and $T_{\rm p}$ denotes the packet transmission time interval. Doing so, it holds that 
\begin{align}
\mathbb{E}[Q]=\frac{1}{a^{2}}\mathbb{E}_{\tilde{Y},\tilde{Z}}\left[e^{a T_{\rm p}\left(\tilde{Y}+\tilde{Z}\right)}-e^{a T_{\rm p}\tilde{Z}}\right]-\frac{T_{\rm p}}{a}\mathbb{E}[\tilde{Y}].
\label{Qk}
\end{align}
Note that $\tilde{Y}$ defines the average number of transmissions between two consecutive (independent) successful receptions. Hence, it follows a geometric probability mass function (PMF), namely ${\rm Pr}[\tilde{Y}=l]=p^{l-1}(1-p),\quad l\geq 1$, where $p$ denotes the packet error probability (PEP). Also, we get
\begin{align}
\mathbb{E}[\tilde{Y}]=\sum^{\infty}_{l=1}l {\rm Pr}[\tilde{Y}=l]=\frac{1}{1-p}.
\label{expYk}
\end{align}
Unlike $\tilde{Y}$, $\tilde{Z}$ stands for the average number of transmissions of a given packet thereby it has a finite range. In fact, $\tilde{Z}=l$ corresponds to a set of $v M+l$ events with $v$ consecutive failures and $l$ retransmission attempts upon successful reception. The PMF of $\tilde{Z}$ yields as \cite[Eq. (14)]{c:GongChenMa18}
\begin{align}
{\rm Pr}[\tilde{Z}=l]&=\sum^{\infty}_{v=0}{\rm Pr}[\tilde{Y}=v M+l]=\frac{1-p}{1-p^{M}}p^{l-1},
\label{pmfZk}
\end{align} 
with $1\leq l\leq M$ and the corresponding expectation on $\tilde{Z}$ stems as
\begin{align}
\mathbb{E}[\tilde{Z}]=\sum^{M}_{l=1}l {\rm Pr}[\tilde{Z}=l]=\frac{1}{1-p}-\frac{M p^{M}}{1-p^{M}}.
\label{expZk}
\end{align} 
For unbounded number of retransmissions (i.e., $M\rightarrow \infty$), then $\mathbb{E}[\tilde{Z}]=\mathbb{E}[\tilde{Y}]=(1-p)^{-1}$. Based on \eqref{Qk}, it holds that
\begin{align}
\nonumber
\mathbb{E}[Q|\tilde{Z}]&=\sum^{\infty}_{l=1}\mathbb{E}[Q|\tilde{Z},\tilde{Y}] {\rm Pr}[\tilde{Y}=l]\\
&=\frac{e^{a T_{\rm p} (Z+1)}-e^{a T_{\rm p} Z}}{a^{2} (1-p e^{a T_{\rm p}})}-\frac{T_{\rm p}}{a (1-p)}.
\label{Qkz}
\end{align}
With the aid of \eqref{pmfZk}, the unconditional expectation on $Q$ is expressed as $\mathbb{E}[Q]=\sum^{M}_{l=1}\mathbb{E}[Q|\tilde{Z}] {\rm Pr}[\tilde{Z}=l]$ yielding
\begin{align}
\mathbb{E}[Q]=\frac{e^{a T_{\rm p}}(e^{a T_{\rm p}}-1) (1-e^{a T_{\rm p} M}p^{M})}{a^{2} (1-p e^{a T_{\rm p}})^{2}(1-p^{M})(1-p)^{-1}}-\frac{T_{\rm p}}{a (1-p)}.
\label{Qkuncond}
\end{align}
Then, using \eqref{expYk} and \eqref{Qkuncond} into \eqref{Qk}, the non-linear average AoI is derived by
\begin{align}
\overline{C}=\frac{e^{a T_{\rm p}}(e^{a T_{\rm p}}-1) (1-e^{a T_{\rm p} M}p^{M})}{a^{2} (1-p e^{a T_{\rm p}})^{2}(1-p^{M})(1-p)^{-2}}-\frac{1}{a}.
\label{avCcl}
\end{align}
Taking the limit $a\rightarrow 0$ and $M\rightarrow \infty$, $\overline{C}$ reduces to the conventional linear average AoI with or without bound on the maximum number of retransmissions, respectively, as $\overline{C}_{a\rightarrow 0}=T_{\rm p}[M-\frac{M}{1-p^{M}}+\frac{3+p}{2 (1-p)}]$ and $\overline{C}_{\begin{subarray}{c}a\rightarrow 0\\M\rightarrow \infty\end{subarray}}=T_{\rm p}[\frac{3+p}{2 (1-p)}]$. In the ideal case of $p\rightarrow 0^{+}$, the latter expression meets a tight lower bound of the average AoI being $T_{\rm p}\times 3/2$.\\
\emph{Remark 1}: From the mode of the proposed operation, a new update message, say $s_{i}$, can be sampled and transmitted only upon the successful reception of $s_{i-1}$. By defining the interarrival packet generation time duration as $X_{i}\triangleq t_{i}-t_{i-1}$, it follows that $\mathbb{E}[X]=\mathbb{E}[Y]$; thereby reflecting on a stop-and-wait update transmission policy. Since each received update packet is as fresh as possible, the considered AoI herein becomes a lower bound to AoI for any system adopting an arbitrary stochastic queuing approach.\\
\emph{Remark 2}: The time lag $\tau$, defined back in the mutual information operator $\mathcal{I}(s_{t};s_{t-\tau})$, between the $i^{\rm th}$ and $(i-1)^{\rm th}$ status updates is given by $\tau_{i}\triangleq t_{i}-t_{i-1}=Z_{i-1}+Y_{i}-Z_{i}$. Specifically, in accordance to the latter remark, the expected time lag equals $\mathbb{E}[Y]$.

An alternative performance metric is the so-called PAoI. PAoI is more tractable than AoI and, most importantly, may be an even more suitable metric in delay-critical applications and/or when a threshold restriction on status aging is required. Based on Fig.~\ref{fig2}, the PAoI of the $i^{\rm th}$ successfully received packet can be modeled by $C^{(i)}_{\rm p}\triangleq a^{-1}(e^{a (Z_{i-1}+Y_{i})}-1)$. Without delving into details and following a similar methodology as for the derivation of average AoI above, the average PAoI can be directly computed as
\begin{align}
\nonumber
&\overline{C}_{\rm p}=a^{-1}\mathbb{E}_{\tilde{Z},\tilde{Y}}\left[\exp\left(a T_{\rm p}(\tilde{Z}_{i-1}+\tilde{Y}_{i})\right)-1\right]\\
\nonumber
&=\frac{1}{a (1-e^{a T_{\rm p}} p)^{2}(1-p^{M})}\bigg\{e^{2 a T_{\rm p}}(1-2 p+p^{M+2})\\
&+2 e^{a T_{\rm p}}p (1-p^{M})-e^{a (M+2) T_{\rm p}}(1-p)^{2} p^{M}+p^{M}-1\bigg\}.
\label{aPAoI}
\end{align}  
In the classical linear AoI case, \eqref{aPAoI} becomes $\overline{C}_{\rm p}^{(a\rightarrow 0)}=T_{\rm p}(\frac{2}{1-p}-\frac{M p^{M}}{1-p^{M}})$, yielding to a tight lower bound of the average PAoI in the extreme (ideal) scenario when $p\rightarrow 0^{+}$ being $2\times T_{\rm p}$.

Until now, all the derived results are conditioned on $p$. We proceed by analyzing the packet error probability of the considered communication system. First, we assume that a packet error occurs only when $\gamma<\gamma_{\rm th}$ for any time instance during an entire packet duration with $\gamma$ representing the instantaneous received signal-to-noise ratio (SNR) and $\gamma_{\rm th}=2^{R}-1$ denoting a certain data rate threshold with $R$ being the target data rate (in bps/Hz). This process is modeled by using the two-state Markov model analyzed in \cite[\S III.B]{j:Fukawa12}. Thereupon, PEP becomes \cite[Eq. (58)]{j:Fukawa12}
\begin{align}
p=1-\exp\left(-\frac{T_{\rm p}{\rm LCR}_{\gamma}(\gamma_{\rm th})}{1-F_{\gamma}(\gamma_{\rm th})}\right)\left[1-F_{\gamma}(\gamma_{\rm th})\right],
\label{PER1}
\end{align}
where ${\rm LCR}_{\gamma}(\cdot)$ and $F_{\gamma}(\cdot)$ represent the level crossing rate (LCR) and cumulative distribution function (CDF) of $\gamma$, respectively. Notably, the defined PEP considers the time variations and correlations of the channel, captured by the second-order LCR statistic. Additionally, \eqref{PER1} can be applied for various ranges of packet length, including the short packet (uncoded) regime used for a vast majority of machine-type applications. More specifically, for spatially independent Rician-faded channels and isotropic scattering, LCR is given by \cite[Eq. (16)]{j:BeaulieuDong2003}
\begin{align}
{\rm LCR}_{\gamma}(\gamma_{\rm th})=\textstyle \frac{\sqrt{2 \pi} f_{D}\left(\frac{\gamma_{\rm th} (K+1)}{\bar{\gamma}}\right)^{\frac{N}{2}}\displaystyle I_{N-1}\textstyle \left(2 \sqrt{\frac{\gamma_{\rm th} (K+1) N K}{\bar{\gamma}}}\right)}{\exp\left(\frac{\gamma_{\rm th}(K+1)}{\bar{\gamma}}+N K\right)(N K)^{\frac{N-1}{2}}},
\label{LCR}
\end{align}
where $K$ is the Rician$-K$ factor (with $K=0$ denoting Rayleigh fading); $\bar{\gamma}$ is the average received SNR; and $f_{D}=u/\lambda$ is the maximum Doppler frequency reflecting on the movement activity of the source node with $u$ and $\lambda$ standing for the relative mobile speed of the source node and carrier wavelength, correspondingly. In the case of Rayleigh fading, \eqref{LCR} becomes ${\rm LCR}_{\gamma}(\gamma_{\rm th})=\sqrt{2 \pi} f_{D}(\frac{\gamma_{\rm th}}{\bar{\gamma}})^{N-\frac{1}{2}}\frac{e^{-\gamma_{\rm th}/\bar{\gamma}}}{(N-1)!}$. In addition, the CDF of $\gamma$ is obtained as \cite[Eq. (22)]{c:KoAbdi00}
\begin{align}
F_{\gamma}(\gamma_{\rm th})=1-Q_{N}\left(\sqrt{2 N K},\sqrt{2 \gamma_{\rm th} (K+1)/\bar{\gamma}}\right),
\label{CDF}
\end{align}
which relaxes to the Erlang distribution for $K=0$ and reduces further to the exponential distribution when $N=1$. Finally, combining \eqref{avCcl} or \eqref{aPAoI} and \eqref{PER1}, the generalized non-linear average AoI or PAoI is captured in a closed-form expression for an arbitrary range of receive antenna arrays, channel fading conditions, and maximum number of packet retransmissions. 

By definition, for a static source node (i.e., when $f_{D}=0$), $p=F_{\gamma}(\gamma_{\rm th})$. Also, the worst-case of PEP (which may also serve as an upper bound; namely, $p_{\rm B}$) is realized when $K=0$ and $N=1$ (i.e., single-antenna receiver and Rayleigh channel fading). In this special case, PEP simplifies to $p_{\rm B}=1-\exp(-\frac{f_{D} T_{\rm p} \sqrt{2 \pi \gamma_{\rm th}\bar{\gamma}}+\gamma_{\rm th}}{\bar{\gamma}})$. Moreover, for quite a small packet length $T_{\rm p}\ll 1$ (which is usually the case in various IoT applications, where $T_{\rm p}\propto 10^{-3}$) and/or asymptotically high SNR regions (when $\bar{\gamma}\rightarrow \infty$), PEP as per \eqref{PER1} approaches $p\approx F_{\gamma}(\gamma_{\rm th})+T_{\rm p} {\rm LCR}_{\gamma}(\gamma_{\rm th})$. From the latter expression, it is obvious that $p\geq F_{\gamma}(\cdot)$ whereas the difference between PEP and CDF of SNR increases for a more intense movement or increasing packet length. Another interesting special case is when the receiver is employed with a massive antenna array (i.e., $N\rightarrow +\infty$). Under Rayleigh channel fading conditions, LCR tends to zero as $N\rightarrow+\infty$ and thereby $p_{(N\rightarrow +\infty)}=F^{(N\rightarrow +\infty)}_{\gamma}(\gamma_{\rm th})$, where $F^{(N\rightarrow +\infty)}_{\gamma}(\cdot)$ is the asymptotic Erlang CDF that can be simplified with the aid of \cite[Eq. (8.11.5)]{b:NIST} as $F^{(N\rightarrow +\infty)}_{\gamma}(\gamma_{\rm th})\approx \frac{(\gamma_{\rm th}/\bar{\gamma})^{N}}{N!}\exp(-\frac{\gamma_{\rm th}}{\bar{\gamma}})$.

Another key performance metric is the energy efficiency (in bits/Hz/Joule), which is defined as 
\begin{align}
\nonumber
{\rm EE}&=\frac{R}{P_{\rm sx}+\mathbb{E}[\tilde{Z}](P_{\rm tx}+N P_{\rm rx})}\\
&=R\left(P_{\rm sx}+\left[\left(\frac{1}{1-p}-\frac{M p^{M}}{1-p^{M}}\right)(P_{\rm tx}+N P_{\rm rx})\right]\right)^{-1},
\label{EE}
\end{align}
where $P_{\rm sx}$, $P_{\rm tx}$ and $P_{\rm rx}$ is the power used for sensing, packet transmission and reception (per antenna), respectively. In the numerator of the first equality on \eqref{EE}, the transmission rate is denoted. In the denominator, the sensing/sampling power, the power used for data communication at the transceiver as well as the number of retransmissions due to erroneous detection are all considered. 

It is noteworthy that $\overline{C}$ and ${\rm EE}$ denote analogous quantities. This holds also for $\overline{C}_{\rm p}$ and ${\rm EE}$. Particularly, as AoI is reduced (i.e., the information updating tends to be more fresh), the energy efficiency is correspondingly reduced since more frequent message transmissions are required. To this end, the AoI/EE and PAoI/EE ratios are introduced, respectively, as 
\begin{align}
\eta \triangleq \overline{C}/{\rm EE}\quad \textrm{and}\quad \eta_{\rm p} \triangleq \overline{C}_{\rm p}/{\rm EE}.
\label{eta}
\end{align}
Minimizing $\eta$ and/or $\eta_{\rm p}$ is quite a challenging and non-convex problem for the general setting of arbitrary Rician faded channels, receive antenna arrays and mobile speed source profiles. It is thus further evaluated in the following section.

\section{Numerical Results and Discussion}
The derived analytical results are verified via numerical validation, whereas they are cross-compared with corresponding Monte-Carlo simulations. Hereinafter, line-curves and circle-marks denote the analytical and simulation results, respectively. Without loss of generality and for ease of presentation, $P_{\rm sx}=P_{\rm tx}=P_{\rm rx}\triangleq P$ is set. Also, unit-valued average channel fading gains are assumed, i.e., $\bar{\gamma}=P/N_{0}$, where $N_{0}$ is the additive white Gaussian noise power. 

\begin{figure}[!t]
\centering
\includegraphics[trim=1.5cm 0.2cm 0.5cm .1cm, clip=true,totalheight=0.35\textheight]{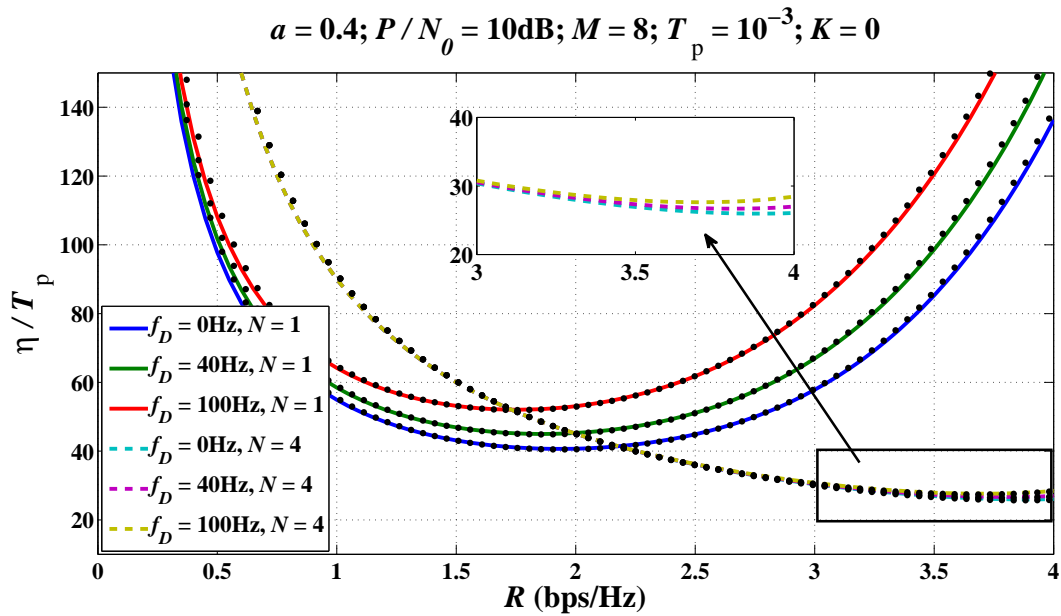}
\caption{Normalized $\eta$ vs. transmission rate $R$ under different receive antenna arrays and source mobile speed.}
\label{fig3}
\end{figure}

\begin{figure}[!t]
\centering
\includegraphics[trim=1.5cm 0.2cm 0.5cm .1cm, clip=true,totalheight=0.35\textheight]{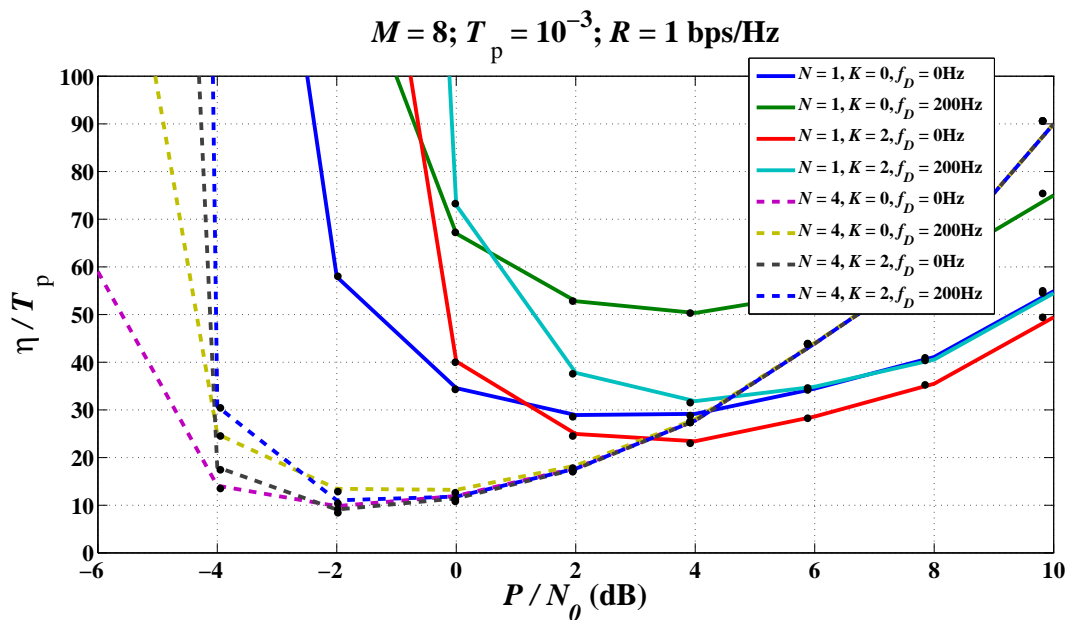}
\caption{Normalized $\eta$ of the linear AoI ($a\rightarrow 0^{+}$) vs. various transmit SNR values under different receive antenna arrays, channel fading conditions and source mobility profile.}
\label{fig4}
\end{figure}

In Fig.~\ref{fig3}, the performance of $\eta$ is depicted where the non-linear AoI case with $a=0.4$ is considered under Rayleigh faded channels. Obviously, extremely low or high data rate values become energy inefficient and/or maximize AoI. Further, the presence of increasing mobility dramatically affects the performance of $\eta$ for the single-antenna deployment. On the other hand, using multiple antennas at the receiver side ($N=4$) greatly enhances the performance in terms of data rate transmission. Insightfully, it seems that the source user mobility marginally affects the performance of $\eta$ for an increasing receive antenna array.  
Quite similar outcomes can be extracted from Fig.~\ref{fig4}, where the standard linear AoI (with $a\rightarrow 0^{+}=10^{-8}$) is considered. In particular, the presence of four receive antennas against the single-antenna scenario reflects on a $6$dB gain on the transmit SNR (on average) regardless of the mobility and/or channel fading conditions. In general, the performance of $\eta$ is being enhanced for higher receive antenna arrays and/or better channel fading conditions (higher Rician$-K$ factor) in terms of energy efficiency. This is a reasonable outcome since PEP is reduced for the latter conditions.

\begin{figure}[!t]
\centering
\includegraphics[trim=1.5cm 0.2cm 0.5cm .1cm, clip=true,totalheight=0.35\textheight]{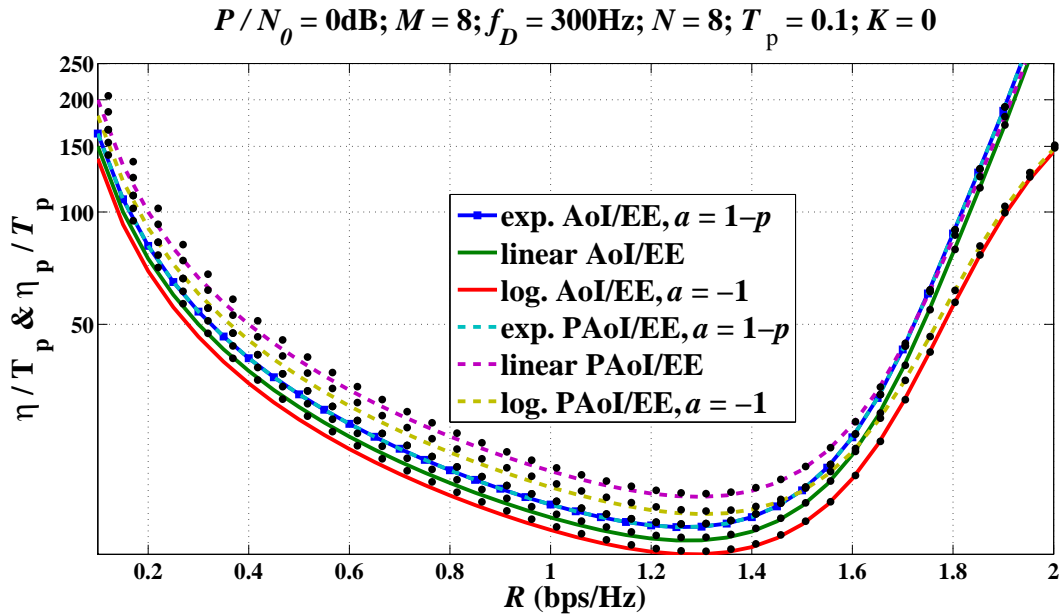}
\caption{Normalized $\eta$ and $\eta_{\rm p}$ vs. transmission rate $R$ under various system setups.}
\label{fig5}
\end{figure}

\begin{figure}[!t]
\centering
\includegraphics[trim=0.5cm 0.0cm 0.2cm .0cm, clip=true,totalheight=0.35\textheight]{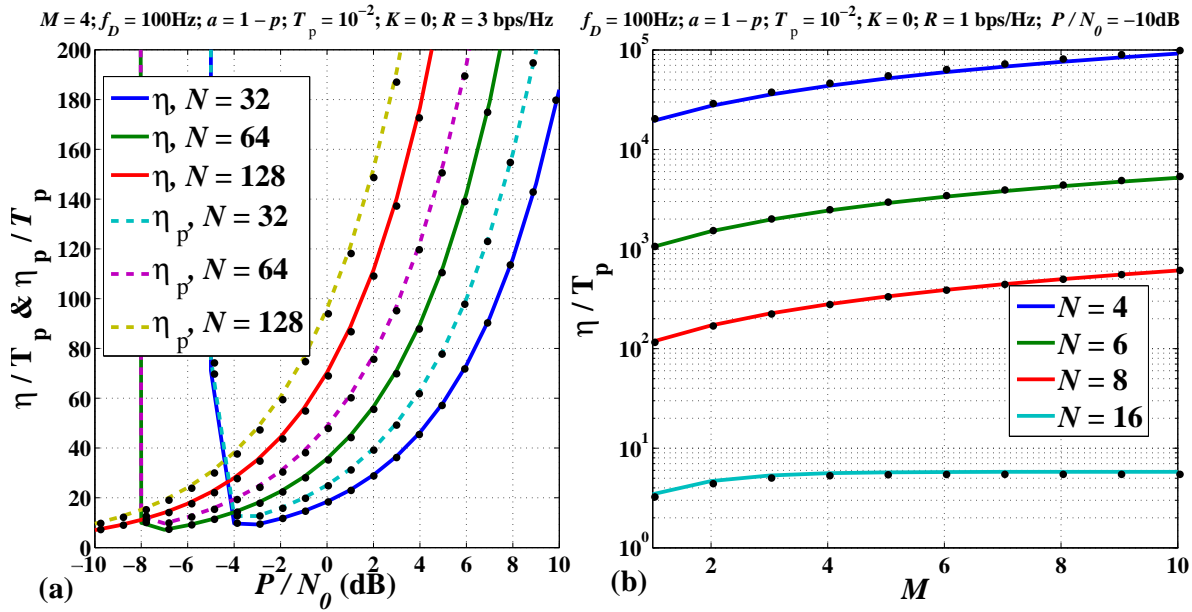}
\caption{(a.) Normalized $\eta$ and $\eta_{\rm p}$ vs. transmit SNR under various large-scale receive antenna arrays; (b.) Normalized $\eta$ vs. $M$ for different receive antenna arrays.}
\label{fig6}
\end{figure}

Both Figs.~\ref{fig3} and~\ref{fig4} above consider the short packet regime (i.e., $T_{\rm p}=10^{-3}$), which is a common case study in various low-latency applications. To better reveal the impact of $a$ parameter on the non-linear AoI, Fig.~\ref{fig5} illustrates the performance of both $\eta$ and $\eta_{\rm p}$ in the case when $N=8$; under a moderately low transmit SNR region, high relative user mobility and longer packet duration being $T_{\rm p}=10^{-1}$. For the exponential AoI scenario, $a=1/\mathbb{E}[X]=1-p$ is set (which equals to the packet generation rate) so as not to overestimate the cost of data aging according to the adopted stop-and-wait transmission policy. As expected, the log-like AoI function provides a reduced $\eta$ and $\eta_{\rm p}$ in the entire range of the provided data rate. This is due to the high correlation between consecutive status update messages which in turn result to a rather low value of information for each new generating sample. On the other hand, the exp-like AoI function presents quite the opposite result. It is also noteworthy that the performance difference between AoI and PAoI tends to vanish for an increasing data rate ($p\rightarrow 1$ as $R$ grows). In this case, $\{\overline{C},\overline{C}_{\rm p}\}\rightarrow +\infty$ yielding convergence between $\eta$ and $\eta_{\rm p}$. Notably, for the considered setup of Fig.~\ref{fig5}, all the AoI/PAoI functions are being minimized in approximately the same range of $R\approx 1.3$bps/Hz.   
In Fig.~\ref{fig6}a, both $\eta$ and $\eta_{\rm p}$ in the large antenna array regime are illustrated. Although PEP tends to considerably be reduced in such a case, the energy efficiency also reduces since an $N-$fold power consumption occurs at the receiver as per \eqref{EE}. Interestingly, it is demonstrated that a higher $N$ is preferable in terms of the minimization of $\eta$ and/or $\eta_{\rm p}$ only in the low transmit SNR regime. Finally, Fig.~\ref{fig6}b presents the performance of $\eta$ under a variable maximum number of allowable retransmissions $M$ in the low SNR regime. Notably, AoI is being increased as $M$ grows, as expected. It is thus preferable not to allow retransmissions (i.e., $M=1$) with regards to the efficiency of both AoI and EE; which is in accordance to other studies, e.g., \cite{j:MangangWang2020}. However, the number of receive antennas plays a key role since there is an enormous performance gap for different $N$ values; this occurs because PEP is considerably reduced for a higher number of receive antennas.  

{\color{black}\section{Conclusions}
The non-linear AoI and PAoI performance metrics were analytically studied under Rician-faded channels, when a single-antenna transmitter communicates directly with a multiple-antenna receiver. The conventional linear (in time) AoI and PAoI arise as special cases of the considered metrics. Both the short- and long-packet regimes were considered, when uncoded transmission is performed. The rather simple yet practically feasible stop-and-wait transmission approach was adopted, which may serve as a performance benchmark for arbitrary queuing models. From the energy efficiency standpoint, it was explicitly demonstrated that the presence of user mobility dramatically affects the system performance. The beneficial role of a multiple-antenna array at the receiver was also revealed. Finally, neither a low nor a relatively high target data rate and/or transmit SNR are preferable; whereas there is an intermediate optimal point that enhances the system performance in terms of both AoI/PAoI and EE.}

\bibliographystyle{IEEEtran}
\bibliography{IEEEabrv,References}

\vfill

\end{document}